\documentclass[useAMS,usenatbib]{mn2e}

\usepackage{graphicx}
\usepackage{color}

\usepackage{gensymb}

\title[Multifrequency observations of SGR~J1745$-$2900]{Simultaneous multifrequency radio
observations of the Galactic Centre magnetar SGR~J1745$-$2900}
\medskip
\author[P.~Torne et al.]{P.~Torne,$^1$\thanks{E-mail: ptorne@mpifr-bonn.mpg.de}
R.~P.~Eatough,$^1$
R.~Karuppusamy,$^1$
M.~Kramer,$^{1,2}$
G.~Paubert,$^3$ \newauthor
B.~Klein,$^{1,4}$ 
G.~Desvignes,$^1$
D.~J.~Champion,$^1$
H.~Wiesemeyer,$^1$
C.~Kramer,$^3$ \newauthor
L.~G.~Spitler,$^1$
C.~Thum,$^3$
R.~G\"{u}sten,$^1$
K.~F.~Schuster,$^5$
and
I.~Cognard$^{6,7}$\\
$^{1}$Max-Planck-Institut f\"{u}r Radioastronomie, Auf dem H\"{u}gel 69, D-53121, Bonn, Germany \\
$^{2}$Jodrell Bank Centre for Astrophysics, School of Physics and Astronomy, The University of Manchester, Manchester M13 9PL, UK\\
$^{3}$Instituto de Radioastronom\'{\i}a Milim\'{e}trica, Avda. Divina Pastora 7, N\'{u}cleo Central, 18012, Granada, Spain\\
$^{4}$Hochschule Bonn-Rhein-Sieg, Grantham-Allee 20, D-53757, Sankt Augustin, Germany\\
$^{5}$Institut de Radioastronomie Millim\'{e}trique, 300 rue de la Piscine, F-38406, Saint Martin d'H\`{e}res, France\\
$^{6}$Laboratoire de Physique et Chimie de l'Environnement et de l'Espace CNRS-Universit{\'e} d'Orl{\'e}ans, F-45071, Orl{\'e}ans, France\\
$^{7}$Station de radioastronomie de Nan{\c c}ay, Observatoire de Paris, CNRS/INSU, F-18330, Nan{\c c}ay, France
}

\begin{document}

\date{Accepted 2015 April 22.  Received 2015 April 16; in original form 2015 March 21}

\pagerange{\pageref{firstpage}--\pageref{lastpage}} \pubyear{2015}

\maketitle

\label{firstpage}

\begin{abstract}
We report on simultaneous observations of the magnetar SGR~J1745$-$2900 at frequencies 
$\nu=\,$2.54 to $\rm{225\,GHz}$ using the \mbox{Nan\c{c}ay$\,$94-m equivalent},
\mbox{Effelsberg$\,$100-m}, and \mbox{IRAM$\,$30-m} radio telescopes. 
We detect  SGR~J1745$-$2900 up to $\rm{225\,GHz}$, the highest radio frequency detection of pulsed emission
from a neutron star to date. 
Strong single pulses are also observed from 4.85 up to $\rm{154\,GHz}$.
At the millimetre band we see significant flux density and spectral index variabilities
on time scales of tens of minutes, plus variability between days at all frequencies.
Additionally, SGR~J1745$-$2900 was observed at a different epoch at frequencies $\nu=\,$296 to $\rm{472\,GHz}$
using the \mbox{APEX$\,$12-m} radio telescope, with no detections.
Over the period $\rm{MJD\,56859.83-56862.93}$ the fitted spectrum yields a spectral index of
$\left<\alpha\right> = -0.4 \pm 0.1$ for a reference flux density $\left< S_{154} \right> = 1.1 \pm 0.2\rm{\,mJy}$
(with $S_{\nu} \propto \nu^{\alpha})$, a flat spectrum alike those of
the other radio-loud magnetars. 
These results show that strongly magnetized neutron stars can be effective radio emitters at frequencies
notably higher to what was previously known 
and that pulsar searches in the Galactic Centre are possible in the millimetre band.

\end{abstract}



\section{Introduction}

Magnetars are thought to be neutron stars with extremely strong magnetic fields \citep{dt92}.
Further work by \citet{td95, td96} theorized how
the decay of the strong magnetic fields in magnetars can produce soft gamma-ray and X-ray emission 
and proposed that Anomalous X-ray Pulsars (AXPs) and Soft Gamma Repeaters (SGRs) could indeed
be magnetars. This identification of AXPs and SGRs as magnetars was validated when pulsations with spin-down 
(a typical characteristic of pulsars) were first detected from one of these objects \citep{kou98}.
At the time of writing, 23 magnetars have been discovered \citep[for a review, see][]{ok14}, and
only four have conclusively shown pulsed radio emission \citep{cam06, cam07a, lev10, eat13, sj13}. 
SGR~J1745$-$2900, the subject of this work, is the last one discovered, and is a compelling object
due to its proximity to Sagittarius A* (Sgr$\,$A*), the supermassive black hole candidate in the
Galactic Centre (GC). Initially discovered at high energies \citep{ken13, mori13, rea13}, pulsations
in the radio band up to $\rm{20\,GHz}$ were also detected with different radio telescopes \citep{eat13, sj13}.

One peculiar characteristic of radio-loud magnetars is that they tend to show shallower spectral
indices than ordinary pulsars.
This property makes magnetars
unique sources to obtain pulsar observations at high radio frequencies, where ordinary pulsars are, with a 
few exceptions \citep{kra97b, morr97, loh08}, usually too faint to be detected and studied. This 
underexplored region of the spectrum of emission is valuable in the study of pulsar
emission physics. For instance some emission models include effects that may only be detectable at high
radio frequencies, like ``coherence breakdown'' between the radio and infrared bands where an
incoherent component of emission becomes dominant \citep{mich82}.
Additionally, unexpected effects have been seen in some of the high frequency
pulsar observations available to date: spectral flattening \citep{kra96}, variability in flux density
\citep{kra97a} or changes in polarization degree \citep{xil96}.

This particular source lies in the vicinity of Sgr$\,$A*. It has been shown
that pulsars closely orbiting the black hole could be used to test General Relativity and alternative
theories of gravity to the highest precision \citep{liu12}. Because the scattering towards the GC
remains not well understood \citep{bow14, spi14}, and scattering may still limit our chances of finding pulsars in the GC, 
SGR~J1745$-$2900 is therefore a useful target for tests of high-frequency observations of the GC.

We present here observations of the magnetar SGR~J1745$-$2900 from 2.54 to $\rm{472\,GHz}$.
Section 2 describes the observations and data reduction. Our results are presented in Section 3. 
Finally, a summary and discussion are given in Section 4.


\section{Observations and data reduction}


\emph{Nan\c cay and Effelsberg:} At the Nan\c cay radio telescope, observations were taken at a
central frequency of $\rm{2.54\,GHz}$ with a bandwidth of $\rm{512\,MHz}$.
Data from dual linear polarizations were coherently dedispersed, folded modulo
the magnetar period and recorded to disc. The observations were flux density calibrated
using the standard {\sc psrchive\footnote{http://psrchive.sourceforge.net/}}
software routines that require the use of a calibrated noise diode
in combination with observations of a calibration source; in this case 3C286 was used.
Observations at
4.85 and $\rm{8.35\,GHz}$ were made at the Effelsberg$\,$100-m
radio telescope using both a digital spectrometer, operating in a pulsar search mode, and
a coherent dedispersion system.
At each frequency dual circular polarizations were recorded over a total bandwidth
of $\rm{500\,MHz}$.
A calibrated noise diode was used to determine the flux density scale by comparing it to NGC7027.
Data were processed with the 
{\sc sigproc\footnote{http://sigproc.sourceforge.net/}} software package.


\emph{\mbox{IRAM$\,$30-m}}: Observations in the millimetre band
were made at the \mbox{IRAM$\,$30-m} radio telescope and used the
Eight MIxer Receiver \citep[EMIR,][]{car12} with the
Broad-Band-Continuum (BBC) backend.  The BBC is connected to four outputs
of EMIR working with dual frequency bands and dual sideband mixers.
Two frequency combinations were used: centred at 87, 101, 138, 154 GHz
(Mode E0/E1) and 87, 101, 209, 225 GHz (mode E0/E2).  The effective
bandwidth of the BBC is estimated to be 24 GHz (6 GHz per frequency
band).  Both horizontal (H) and vertical (V) linear polarizations were
recorded. We noted a systematic lower sensitivity in the H
polarization channel, the cause of which we were not able to identify.
A special set-up of the backend enabled fast time sampling of the
continuum signal at $\rm{1\,ms}$.
The total variations of the system temperature ($T_{\rm{sys}}$) during the four days
were between 77$-$183$\rm{\,K}$ for the $\rm{87\,GHz}$ band,
84$-$208$\rm{\,K}$ for the $\rm{101\,GHz}$ band, 87$-$257$\rm{\,K}$ for $\rm{138\,GHz}$ band, 
98$-$327$\rm{\,K}$ for the $\rm{154\,GHz}$ band, 478$-$571$\rm{\,K}$ for the $\rm{209\,GHz}$ band and
520$-$621$\rm{\,K}$ for the $\rm{225\,GHz}$ band. 
Temporal fluctuations in the atmospheric water vapour are responsible for the high variations of $T_{\rm{sys}}$.
We observed in ``Total Power'' mode, i.e. always on source. This observing mode has
the disadvantage that atmospheric and receiver gain variations are reflected in the data as variations of
the mean count level. 
Self-produced instrumental interference was present in the data, in particular at harmonics
of $\rm{1\,Hz}$, most likely produced by the cryogenerator.

The data reduction was as follows.
The dispersion delay due to the interstellar medium, even at the large $\rm{DM\,=\,1778\,cm^{-3}\,pc}$ of the source,
is only $\rm{\Delta t_{DM}\,\simeq\,0.9\,ms}$ across the observing frequency range. 
This delay is smaller than the sampling time of $\rm{1\,ms}$ and therefore negligible.
The variations in the time series and interference needed to be removed because they contaminated significantly the baseline
of the folded pulsar profile, reducing our sensitivity to the pulsed emission.
Our solution consisted of using a sliding window of width $\rm{3\,s}$ where a sine wave of $\rm{1\,Hz}$ and baseline were
fit to the data, subtracting the central $\rm{1\,s}$ of the fit from the raw data. 
This reduced the variations enough to produce a nearly flat-baseline folded profile.
For 209 and \rm{225\,GHz}, this method was less effective than a running mean filtering, that was applied
twice to the time series with windows of 10 and $\rm{0.4\,s}$, respectively. Both filtering methods subtract the 
continuum emission contribution from the GC.
Finally, the cleaned time series were folded with the topocentric period of SGR~J1745$-$2900, calculated from a
recent ephemeris from a timing analysis performed at lower frequencies.
The flux density calculations were done using interspersed hot-cold load calibration measurements 
at the beginning and end of each observation scan, providing the required quantities to convert counts to antenna temperature Ta*.
The methodology follows \citet{ckra97}. The conversion factor from Ta* to flux density $(S/\rm{Ta*})$ 
was obtained from the observatory efficiency tables\footnote{http://www.iram.es/IRAMES/mainWiki/Iram30mEfficiencies}. 
Pointing corrections were also applied between scans.
In addition, due to the low telescope elevations during the observations, an 
elevation-dependent telescope-gain correction was applied \citep{pen12}.
Our flux density calibration method was verified with pointings on the planets Mars and Uranus.


\emph{APEX}: Additional millimetre and submillimetre data were taken with the APEX radio telescope.
The system used was the FLASH$^+$ receiver with a special version of the continuum Pocket BackEnd (PBE).
FLASH$^+$ worked in dual sideband mode providing simultaneously four bands centred at 296, 308,
460, and $\rm{472\,GHz}$. The total bandwidth was 16 GHz (4 GHz per frequency band),
splitting and recording one polarization at 296 and $\rm{308\,GHz}$ and the other at 460 and $\rm{472\,GHz}$.
The sampling time of the continuum signal was $\rm{0.5\,ms}$.
SGR~J1745$-$2900 was observed on 2014 August 24 for $\rm{60\,min}$.
$T_{\rm{sys}}$ varied during the observation between 119$-$130$\rm{\,K}$ for the $\rm{296\,GHz}$ band,
129$-$143$\rm{\,K}$ for the $\rm{308\,GHz}$ band, 589$-$811$\rm{\,K}$ for the $\rm{460\,GHz}$ band,
and 471$-$3414$\rm{\,K}$ for the $\rm{472\,GHz}$ band.
The atmospheric water vapour content is responsible for the high $T_{\rm{sys}}$ at 460 and $\rm{472\,GHz}$.
Again, ``Total Power'' measurements were done. We applied the two-running-mean method with 10 and $\rm{0.4\,s}$
windows to reduce the variations in the time series and subtract the continuum contributions of the atmosphere and
GC. The cleaned
time series were then folded using the topocentric period of SGR~J1745$-$2900 computed from our ephemeris.
For the non-detection flux density limits, the radiometer equation with a signal-to-noise threshold of 5, duty cycle of
0.075, and telescope gain from the observatory efficiency tables\footnote{http://www.apex-telescope.org/telescope/efficiency/index.php}
is used \citep[see e.g.][]{lorkra05}. Factors to account for the transmissivity of the atmosphere are also included.

Within the constraints of the source visibility the observations from \mbox{IRAM$\,$30-m}, Effelsberg and Nan\c{c}ay were
simultaneous, overlapping a total of $\rm{203\,min}$ on the dates 2014 July 22$-$24. Table 1 summarizes the observations.

\begin{table}
  \centering
  \caption{Summary of the observations of SGR~J1745$-$2900 with measured flux densities and spectral indices. 
  For July 21 the relative error bars are too large for a meaningful spectral index fit.
  Note that the APEX observations are one month later from the rest.
  In the observatory column, (Obs.) I corresponds to \mbox{IRAM$\,$30-m}, N to Nan\c{c}ay,
  E to Effelsberg and A to APEX. Errors in the last digits are shown in parenthesis.
  }
  \label{table:fluxes}
  \begin{tabular}{@{}cccccc@{}}
  \hline
   Date				& Obs.	& $\nu$	& $S$		  & $\alpha$	& $\rm{T_{obs} }$\\
				&	& (GHz) & (mJy)		  &		& (h)\\
 \hline
\rule{0pt}{3ex} 2014 Jul 21	& I	& 87    & $0.1(4)$ & $-$ 		& 2.0\\
				& 	& 101   & $0.6(5)$ & 	 		& 2.0\\
				& 	& 138   & $0.2(3)$ & 	 		& 2.0\\
				& 	& 154   & $0.3(4)$ & 	 		& 2.0\\

\rule{0pt}{3ex} 2014 Jul 22	& N	& 2.54	& $10.5(11)$ & $-0.3(1)$	& 1.3\\
				& E	& 4.85  & $3.9(3)$ & 			& 1.0\\
				& 	& 8.35  & $3.3(3)$ & 			& 0.8\\
				& I	& 87    & $1.5(3)$ & 			& 4.0\\
				& 	& 101 	& $1.8(3)$ & 			& 4.0\\
				& 	& 138 	& $2.2(3)$ & 			& 4.0\\
				& 	& 154 	& $1.5(3)$ & 			& 4.0\\
				
\rule{0pt}{3ex} 2014 Jul 23 	& N	& 2.54	& $11.0(11)$ & $-0.4(1)$	& 1.3\\
				& E	& 4.85  & $4.0(3)$ & 			& 1.1\\
				& 	& 8.35  & $3.3(3)$ & 			& 1.1\\
				& I	& 87  	& $1.8(4)$ & 			& 2.4\\
				& 	& 101 	& $1.7(5)$ & 			& 2.4\\
				& 	& 138 	& $1.7(3)$ & 			& 2.4\\
				& 	& 154 	& $1.9(4)$ & 			& 2.4\\

\rule{0pt}{3ex} 2014 Jul 24	& N	& 2.54	& $3.8(4)$ & $-0.2(1)$		& 1.3\\
				& E	& 4.85  & $2.2(3)$ & 			& 1.1\\
				& 	& 8.35  & $4.2(4)$ & 			& 1.1\\
				& I	& 87  	& $2.6(4)$ & 			& 2.0\\
				& 	& 101 	& $2.1(4)$ & 			& 2.0\\
				& 	& 138 	& $0.8(4)$ & 			& 0.8\\
				& 	& 154 	& $0.6(5)$ & 			& 0.8\\
				& 	& 209 	& $1.6(11)$ & 			& 1.2\\
				& 	& 225 	& $1.0(7)$ &  			& 1.2\\

\rule{0pt}{3ex} 2014 Aug 24	& A	& 296 & $< 2.1 $  & $-$			& 1.0\\
				& 	& 308 & $< 2.2 $  &  			& 1.0\\
				& 	& 460 & $< 29.9 $ &  			& 1.0\\
				& 	& 472 & $< 52.2 $ &  			& 1.0\\
				
\rule{0pt}{3ex} Total average	& N	& 2.54  & $8.4(5)$ & $ -0.4(1) $	& 3.9\\
				& E	& 4.85 & $3.4(2)$ & 			& 3.1\\
				&	& 8.35 & $3.6(2)$ & 			& 3.0\\
				& I	& 87  & $1.5(2)$ &  			& 10.4\\ 
				& 	& 101 & $1.6(2)$ &  			& 10.4\\
				& 	& 138 & $1.5(2)$ &  			& 9.2\\
				& 	& 154 & $1.1(2)$ &  			& 9.2\\
				& 	& 209 & $1.6(11)$ &  			& 1.2\\
				& 	& 225 & $1.0(7)$ &  			& 1.2\\

 \hline
\end{tabular}
\end{table}


\section{Results}

\begin{figure}
    \begin{center}
    \includegraphics[width=0.95\linewidth]{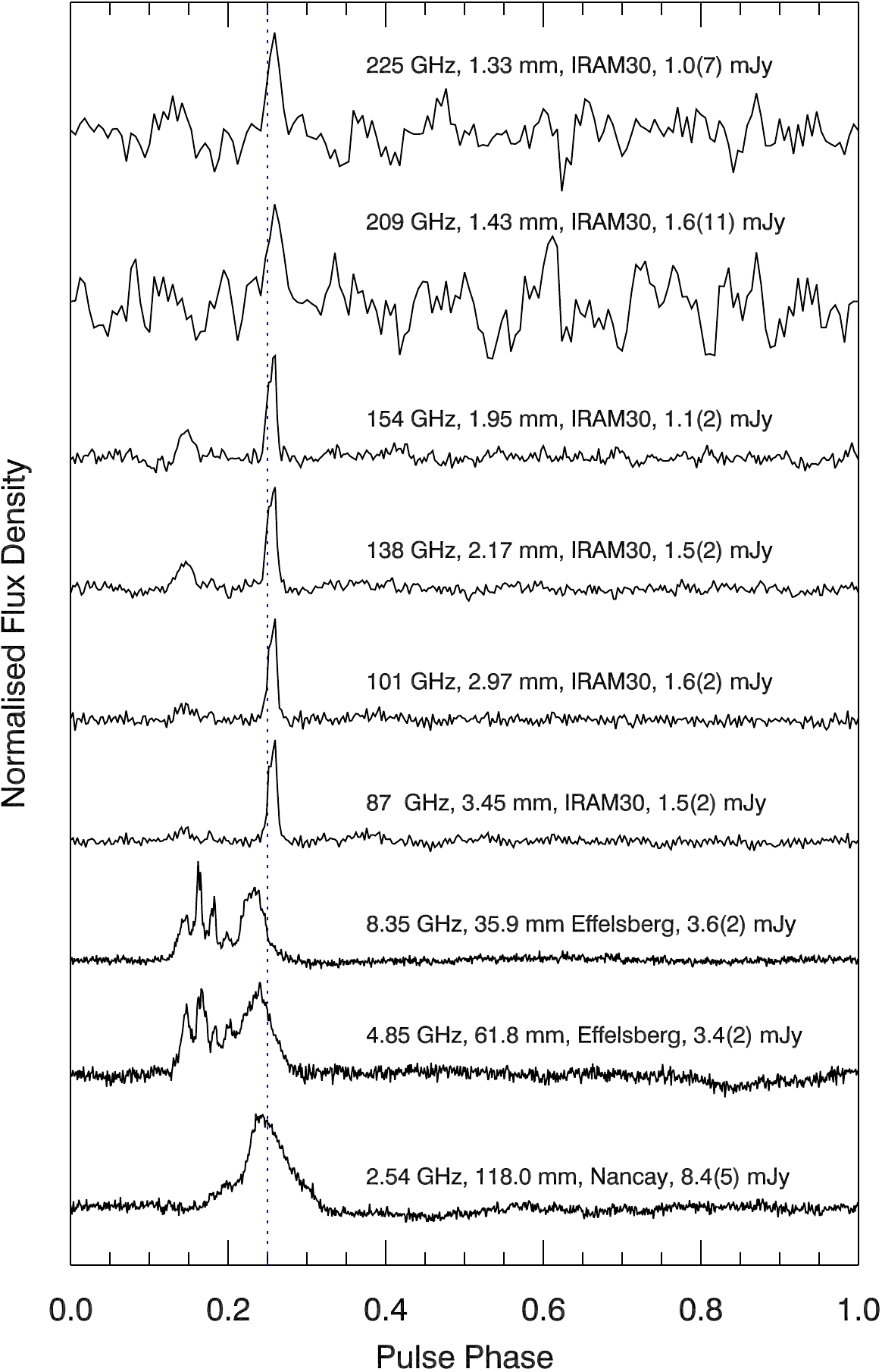}
    \caption{Average pulse profiles of SGR~J1745$-$2900 at
      2.54, 4.85, 8.35, 87, 101, 138, 154, 209 and 225$\,$GHz. All
      profiles are generated from the sum of the observations
      presented here and are aligned to the same reference phase
      obtained from our timing ephemeris (dotted line).
      }
    \label{fig:profiles}
    \end{center}
\end{figure}

The magnetar was clearly detected up to $\rm{154\,GHz}$ and more
weakly detected at 209 and $\rm{225\,GHz}$ with clear peaks at the
expected rotational phase. At APEX, 296 to $\rm{472\,GHz}$, no
  detections were made.  Fig.~\ref{fig:profiles} shows the detected
average profiles at each frequency. The profiles are generally
multicomponent, with a shape that varied significantly from day to
day at the lowest frequencies 2.54 to $\rm{8.35\,GHz}$.  At the
millimetre band the profile was more stable, showing a main peak and a
precursor (pulse phase $\simeq\,0.15$ in Fig.~1). The precursor seems
to consist of different narrow components that appeared and
disappeared randomly between observations, and was on average stronger
at the highest frequencies.

\begin{figure}
    \begin{center}
    \includegraphics[width=0.96\linewidth]{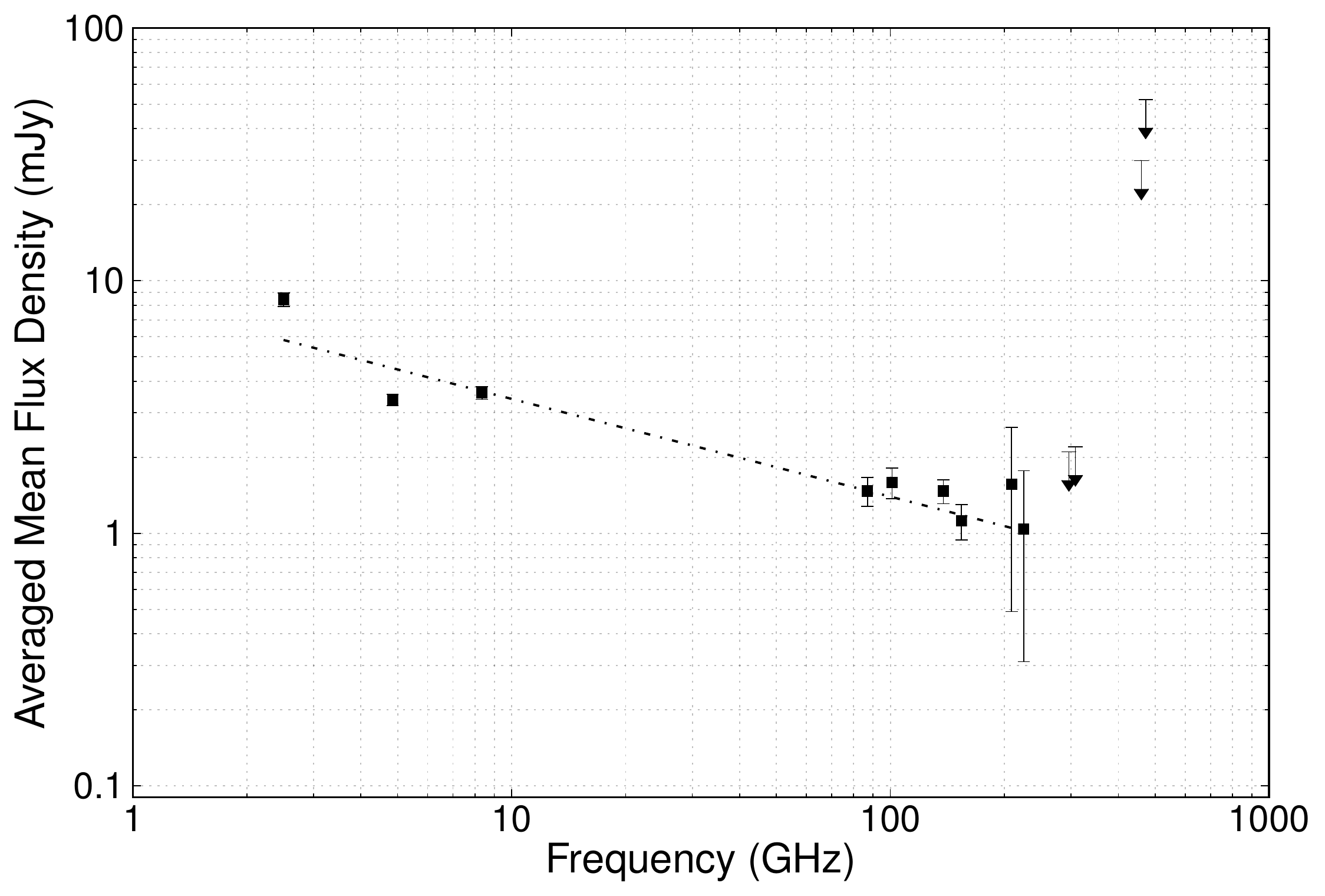}
    \caption{Total averaged mean flux densities. The dash-dotted line shows the spectral index fit.
    The mean spectral index obtained for SGR~J1745$-$2900 is $\rm{\left<\alpha\right>\,=\,-0.4\,\pm\,0.1}$,
    showing the unusual flatter spectrum common to magnetars in the radio band. 
    Due to the variability of the source the APEX data (taken approximately one month later) were not used to 
    prove or disprove the continuation of the power law beyond $\rm{230\,GHz}$.}
    \label{fig:spectrum}
    \end{center}
\end{figure}

In Table~1 we present the equivalent continuum flux density (also known as the ``mean flux density'') 
from all observations. The errors were estimated taking into account the uncertainties in the calibration
processes. We observed significant variations in flux density on short time-scales (of the order of tens of
minutes) in the millimetre data and day-to-day variability at all frequencies.
Fig.~\ref{fig:spectrum} shows the total averaged mean flux density per frequency band and a power-law fit to the
data spanning more than $\rm{220\,GHz}$. Over the period $\rm{MJD\,56859.83-56862.93}$ the fitted spectrum yields a spectral index of
$\left<\alpha\right> = -0.4 \pm 0.1$ for a flux density at $\rm{154\,GHz}$ $\left< S_{154} \right> = 1.1 \pm 0.2\rm{\,mJy}$.

\begin{figure}
    \begin{center}
    \includegraphics[width=\linewidth]{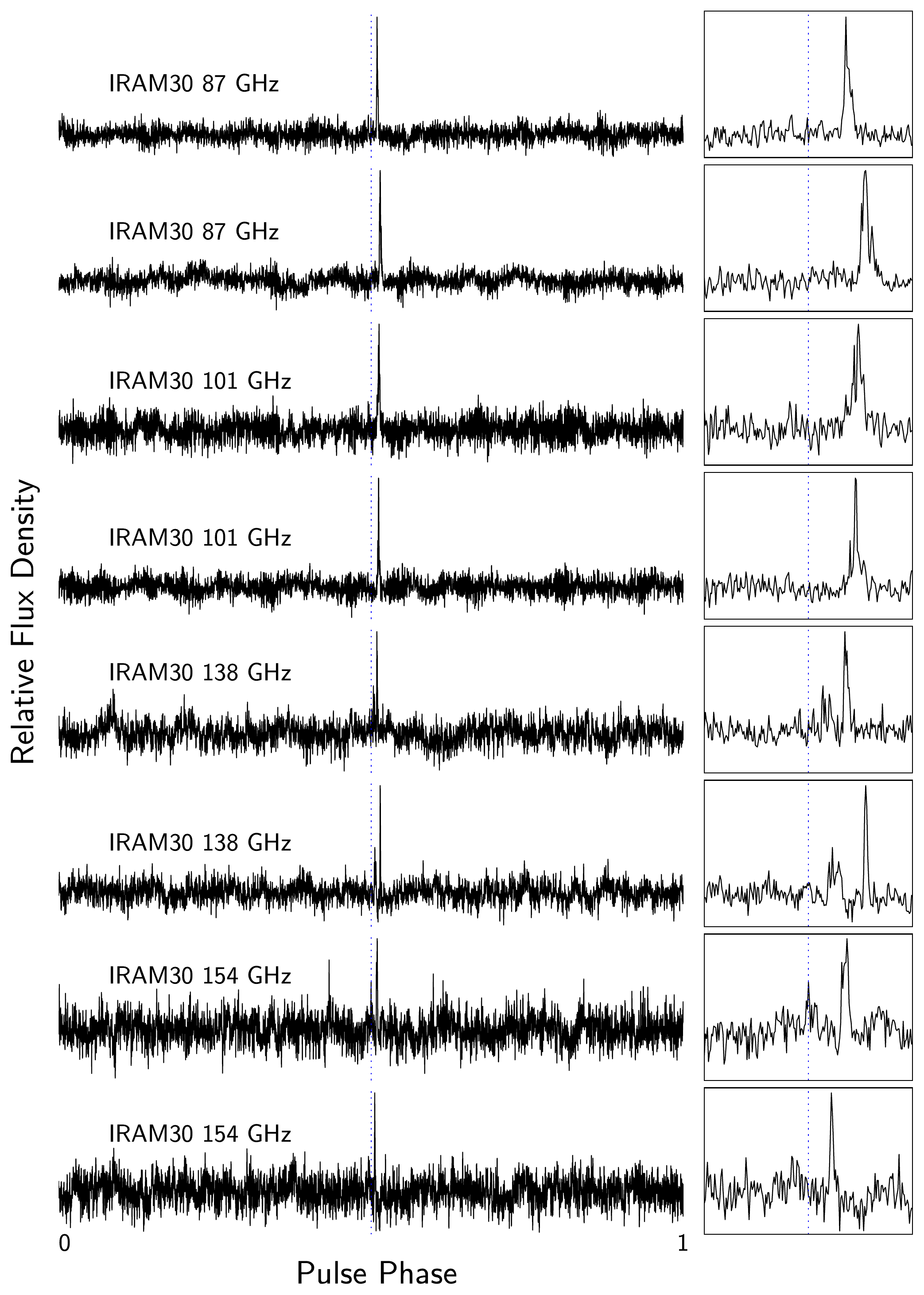}
    \caption{Selection of strong single pulses from our data set in the millimetre band. 
    The time resolution is $\rm{1\,ms}$. The panels on the left show one rotation of the neutron star.
    Panels on the right show a zoom to a window of $\rm{0.2\,s}$ around the reference phase computed from
    the ephemeris (dotted line).
    The width, intensity and morphology of the single pulses are diverse.
    The features visible in the off-pulse region are due to residual interference.
    }
    \label{fig:SPs}
    \end{center}
\end{figure}

We also detected numerous single pulses at all frequencies between 4.85 and $\rm{154\,GHz}$ (at $\rm{2.54\,GHz}$ 
the delivered folded data did not allow to search for single pulses).
At the millimetre band,
the peak flux density of the strongest pulses reached $\rm{19\,Jy}$ at $\rm{101\,GHz}$, with a pulse width
of $\rm{1\,ms}$. Following \citet{lorkra05}, this is equivalent to a brightness temperature $T_{\rm{B}} > 10^{23}\,\rm{K}$,
for a distance to the GC of $\rm{8.3\,kpc}$ \citep{gill09}. 
At $\rm{154\,GHz}$, the strongest pulse reached a peak of $\rm{8\,Jy}$ with a width of $\rm{1\,ms}$ and 
$T_{\rm{B}} \simeq 2.3\cdot10^{22}\,\rm{K}$. The majority of the single pulses detected showed $T_{\rm{B}}$ of the order of
$10^{22}\,\rm{K}$.
The values calculated here are lower limits because we are limited by the coarse time resolution of the data.
Fig.~\ref{fig:SPs} shows a selection of strong single pulses from our data set at the highest frequencies. 
A detailed analysis of the single pulses from SGR~J1745$-$2900 will be presented elsewhere.

Possible reasons for the non-detections at APEX include the
  comparably lower sensitivity of the observations, the faintness of
  the source at these frequencies, and the random variability of the
  source.  We also noted strong, periodic self-produced instrumental
  interference overlapping the magnetar spin frequency that could have
  decreased our sensitivity even further.  The polarization splitting
  of the APEX receiver (see \S2), if the magnetar is highly linearly
  polarized at these frequencies, could have also led to a loss in
  sensitivity if the radiation was misaligned with the feed at the
  lower bands, 296 and $\rm{308\,GHz}$, that were much more sensitive.
  Because of these uncertainties, we cannot conclusively rule out the
  emission of SGR~J1745$-$2900 between 296 and $\rm{472\,GHz}$, nor a
  lack of it.


\section{Summary and Discussion}

\parskip 0pt

The detections of SGR~J1745$-$2900 up to $\rm{225\,GHz}$
and its single pulses up to $\rm{154\,GHz}$ set new records for the
detection of pulsed emission from a neutron star in the radio band.
The high measured flux density of SGR~J1745$-$2900 at the
millimetre band between 87 and $\rm{225\,GHz}$, $\rm{\left<
  \emph{S}_{mm} \right> = 1.4\,\pm\,0.2\,mJy}$, and the numerous
strong single-pulses detected up to $\rm{154\,GHz}$ show that
SGR~J1745$-$2900 can be, at least during some periods of activity, an
efficient producer of radiation at very high radio frequencies.  In
combination with the work by \citet{cam07b}, our detections suggest
that emission at very high radio frequencies from radio-loud magnetars
might be a frequent characteristic of these objects.

The origin of the radio emission from magnetars is still unclear. The
case of magnetars is perhaps more complex than for ordinary pulsars,
which are thought to be powered entirely by their loss of rotational
kinetic energy, (also called ``spin-down luminosity'') $\dot{E}$.
Some magnetars show luminosities at high energies that are much larger
than $\dot{E}$ \citep[see, e.g., Tables~2 and 7 in][]{ok14}, requiring
additional sources to power their electromagnetic emission. Big
efforts have been made trying to model processes that describe the
emission from magnetars \citep{td95, belo07, belo09, belo13}, but the
puzzle is not yet solved. From our observations, we can infer some
properties of the radio emission from SGR~J1745$-$2900. For the high
brightness temperatures of the single pulses, $T_{\rm{B}} \sim [10^{22} -
  10^{23}]\,\rm{K}$, a coherent mechanism for the radio emission is
required. Following \citet{lorkra05}, its radio luminosity is
estimated to be $R_{\rm{lum}} \sim 10^{29}\,\rm{erg\,s^{-1}}$,
assuming a duty cycle of 0.15 and an emission cone opening angle of
$\rm{4\degree}$. Its spin-down luminosity is $\dot{E} \sim
[10^{33}-10^{34}]\,\rm{erg\,s^{-1}}$, depending on what timing
solution is used \citep[our own, or e.g.][]{kas14, lyn14}.  Therefore,
$R_{\rm{lum}}/\dot{E} \sim [10^{-4}-10^{-5}]$. This shows that the
radio emission could be powered by its rotational kinetic energy;
although this does not mean that this is the source or rule out that
other mechanisms are involved.
SGR~J1745$-$2900 shows common characteristics with
ordinary pulsars, and remarkable differences, like it is for the
other three known radio magnetars \citep{kra07, cam08, laz08, lev12}.
The reason for this, as well as the details of the radio emission
from pulsars in general, is still unknown.

Finally, the detection of SGR~J1745$-$2900 up to 225$\,$GHz at a
  distance of 8.3$\,$kpc shows that it is, in principle, possible to
  search for pulsars in the GC at these
  frequencies. A key advantage of pulsar searches at millimetre
  wavelengths is that any deleterious effects caused by the GC
  interstellar medium (i.e. pulse scattering and dispersion) can be
  fully neglected. Along the same lines, any deleterious effects caused 
  by the intergalactic medium, which are not well known, could also be
  neglected or will have a minimal impact. It might
  therefore also be an option to search for shallow-spectrum pulsars
  in nearby galaxies through single-pulse emission.
  At 1.4$\,$GHz, the magnetar is more luminous
  than 97 per cent of all known pulsars, and it is by far the most
  luminous in the millimetre band. This fact alone suggests a larger
  less-luminous population that can be detected with increased
  sensitivity at the millimetre band.  In any case, pulsar searches
  with more sensitive millimetre telescopes, such as ALMA, are
  promising.

\vspace*{-2mm}

\section*{Acknowledgements}

We are grateful to the referee S. Johnston for very useful
  comments, to A. Jessner and D. Riquelme for discussions,
  to P. Lazarus for sharing code, and to C. Ng for
  reading the manuscript.  Partly based on observations with the 100-m
  telescope of the MPIfR at Effelsberg.  The Nan{\c c}ay radio
  observatory is operated by the Paris Observatory, associated to the
  French CNRS.  IRAM is supported by INSU/CNRS (France), MPG (Germany)
  and IGN (Spain).  APEX is a collaboration between the MPIfR, the
  European Southern Observatory, and the Onsala Space Observatory. PT
  is supported for this research through a stipend from the 
  International Max Planck Research School (IMPRS) for
  Astronomy and Astrophysics at the Universities of Bonn and Cologne.

\bsp
\label{lastpage}


\vspace*{-3mm}


\begin{thebibliography}{99}
\raggedright

\bibitem[\protect\citeauthoryear{Beloborodov}{2009}]{belo09} Beloborodov A.~M., 2009,
 ApJ, 703, 1044
\bibitem[\protect\citeauthoryear{Beloborodov}{2013}]{belo13} Beloborodov A.~M., 2013, ApJ, 777, 114
\bibitem[\protect\citeauthoryear{Beloborodov \& Thompson}{2007}]{belo07} Beloborodov A.~M., Thompson C.,
 2007, ApJ, 657, 967
\bibitem[\protect\citeauthoryear{Bower et al.}{2014}]{bow14} Bower G. C. et al.,
 2014, ApJ, 780, L2
\bibitem[\protect\citeauthoryear{Camilo et al.}{2006}]{cam06} Camilo F.,
 Ransom S.~M., Halpern J.~P., Reynolds J., Helfand D.~J., Zimmerman N., Sarkissian J.,
 2006, Nature, 442, 892
\bibitem[\protect\citeauthoryear{Camilo et al.}{2007a}]{cam07a} Camilo F., Ransom S.~M.,
 Halpern J.~P., Reynolds J., 2007a, ApJ, 666, L93
\bibitem[\protect\citeauthoryear{Camilo et al.}{2007b}]{cam07b} Camilo F. et al., 2007b, ApJ,
 669, 561
\bibitem[\protect\citeauthoryear{Camilo et al.}{2008}]{cam08} Camilo F., Reynolds J., Johnston S.,
 Halpern J.~P., Ransom S.~M., 2008, ApJ, 679, 681
\bibitem[\protect\citeauthoryear{Carter et al.}{2012}]{car12} Carter M. et al., 2012, A\&A,
 538, A89
\bibitem[\protect\citeauthoryear{Duncan \& Thompson}{1992}]{dt92} Duncan R.~C.,
 Thompson C., 1992, ApJ, 392, L9
\bibitem[\protect\citeauthoryear{Eatough et al.}{2013}]{eat13} Eatough R.~P. et al., 2013, Nature,
 501, 391
\bibitem[\protect\citeauthoryear{Gillessen et al}{2009}]{gill09} Gillessen S., Eisenhauer F., Trippe S.,
 Alexander T., Genzel R., Martins F., Ott T., 2009, ApJ, 692, 1075
\bibitem[\protect\citeauthoryear{Kaspi et al.}{2014}]{kas14} Kaspi V.~M. et al., 2014, ApJ, 786, 84
\bibitem[\protect\citeauthoryear{Kennea et al.}{2013}]{ken13} Kennea J.~A. et al., 2013, ApJ,
 770, L24
\bibitem[\protect\citeauthoryear{Kouveliotou et al.}{1998}]{kou98} Kouveliotou C. et al., 
1998, Nature, 393, 235
\bibitem[\protect\citeauthoryear{Kramer}{1997}]{ckra97} Kramer C., 1997, IRAM Internal Technical Memo
\bibitem[\protect\citeauthoryear{Kramer et al.}{1996}]{kra96} Kramer M., Xilouris K.~M.,
 Jessner A., Wielebinski R., Timofeev M., 1996, A\&A, 306, 867
\bibitem[\protect\citeauthoryear{Kramer et al.}{1997a}]{kra97a} Kramer M., Xilouris K.~M.,
 Rickett B., 1997a, A\&A, 321, 513
\bibitem[\protect\citeauthoryear{Kramer et al.}{1997b}]{kra97b} Kramer M., Jessner A., Doroshenko O.,
 Wielebinski R., 1997b, ApJ, 488, 364
\bibitem[\protect\citeauthoryear{Kramer et al.}{2007}]{kra07} Kramer M., Stappers B.~W., Jessner A.,
 Lyne A.~G., Jordan C.~A., 2007, MNRAS, 377, 107
\bibitem[\protect\citeauthoryear{Lazaridis et al.}{2008}]{laz08} Lazaridis K., Jessner A.,
 Kramer M., Stappers B.~W., Lyne A.~G., Jordan C.~A., Serylak M., Zensus J.~A., 2008, MNRAS,
 390, 839
\bibitem[\protect\citeauthoryear{Levin et al.}{2010}]{lev10} Levin L. et al., 2010, ApJ,
 721, L33
\bibitem[\protect\citeauthoryear{Levin et al.}{2012}]{lev12} Levin L. et al., 2012, MNRAS,
 422, 2489
\bibitem[\protect\citeauthoryear{Liu et al.}{2012}]{liu12} Liu K., Wex N., Kramer M., Cordes J.~M., 
 Lazio T.~J.~W., 2012, ApJ, 747, 1 
\bibitem[\protect\citeauthoryear{L\"{o}hmer et al.}{2008}]{loh08} L{\"o}hmer O., Jessner A., Kramer M.,
 Wielebinski R., Maron O., 2008, A\&A, 480, 623
\bibitem[\protect\citeauthoryear{Lorimer \& Kramer}{2005}]{lorkra05} Lorimer D.~R., Kramer M., 2005,
 Handbook of Pulsar Astronomy, Cambridge University Press, Cambridge, UK
\bibitem[\protect\citeauthoryear{Lynch et al.}{2014}]{lyn14} Lynch R.~S., Archibald R.~F., 
 Kaspi V.~M., Scholz P., 2014, preprint (arXiv:1412.0610)
\bibitem[\protect\citeauthoryear{Michel}{1982}]{mich82} Michel F.~C., 1982, Reviews of Modern Physics,
 54, 1
\bibitem[\protect\citeauthoryear{Mori et al.}{2013}]{mori13} Mori K. et al., 2013, ApJ, 770,
 L23
\bibitem[\protect\citeauthoryear{Morris et al.}{1997}]{morr97} Morris D. et al., 1997, A\&A, 322, L17
\bibitem[\protect\citeauthoryear{Olausen \& Kaspi}{2014}]{ok14} Olausen S.~A.,
 Kaspi V.~M., 2014, ApJS, 212, 6
\bibitem[\protect\citeauthoryear{Pe\~{n}alver}{2012}]{pen12} Pe\~{n}alver J., 2012, 
 IRAM Internal Technical Memo 
\bibitem[\protect\citeauthoryear{Rea et al.}{2013}]{rea13} Rea N. et al., 2013, ApJ, 775, L34
\bibitem[\protect\citeauthoryear{Shannon \& Johnston}{2013}]{sj13} Shannon R.~M., Johnston S.,
 2013, MNRAS, 435, L29
\bibitem[\protect\citeauthoryear{Spitler et al}{2014}]{spi14} Spitler L.~G. et al., 2014, ApJ,
 780, L3
 \bibitem[\protect\citeauthoryear{Thompson \& Duncan}{1995}]{td95} Thompson C.,
 Duncan R.~C., 1995, MNRAS, 275, 255
\bibitem[\protect\citeauthoryear{Thompson \& Duncan}{1996}]{td96} Thompson C.,
 Duncan R.~C., 1996, ApJ, 473, 322
\bibitem[\protect\citeauthoryear{Xilouris et al.}{1996}]{xil96} Xilouris K.~M., Kramer M.,
 Jessner A., Wielebinski R., Timofeev M., 1996, A\&A, 309, 481

\end{thebibliography}
\end{document}